\documentclass[aps,prl,twocolumn,groupedaddress,amsmath,amssymb,floatfix]{revtex4}
\pdfoutput=1

\usepackage{graphicx}
\usepackage{tabularx}
\usepackage[usenames]{color}
\usepackage[usenames,x11names]{xcolor}

\definecolor{tsampgreen}{rgb}{0,0.6,0}

\begin{document}

\title{Exponentially Fragile $\mathcal{PT}$-Symmetry in Lattices with Localized Eigenmodes}

\author{Oliver Bendix$^{1}$}
\author{Ragnar Fleischmann$^{1}$}
\author{Tsampikos Kottos$^{2}$}
\author{Boris Shapiro$^{3}$}
\affiliation{$^1$MPI for Dynamics and Selforganization, Bunsenstrasse 10, Goettingen, Germany}
\affiliation{$^2$Wesleyan University, Middletown, Connecticut 06459}
\affiliation{$^3$Technion - Israel Institute of Technology, Technion City, Haifa 32000, Israel}

\date{\today}

\begin{abstract}
We study the effect of localized modes in lattices of size $N$ with parity-time (${\cal PT}$) symmetry.
Such modes are arranged in pairs of quasi-degenerate levels with splitting $\delta \sim \exp^{ -N/\xi}$ 
where $\xi$ is their localization length. The level "evolution" with respect to the ${\cal PT}$ breaking 
parameter $\gamma$ shows a cascade of bifurcations during which a pair of real levels becomes complex. 
The spontaneous ${\cal PT}$ symmetry breaking occurs at $\gamma_{\cal PT} \sim {\rm min} \{\delta\}$, 
thus resulting in an exponentially narrow exact ${\cal PT}$ phase. As $N/\xi$ decreases, it becomes more 
robust with $\gamma_{\cal PT}\sim 1/N^2$ and the distribution ${\cal P} (\gamma_{\cal PT})$ changes 
from log-normal to semi-Gaussian. Our theory can be tested in the frame of optical lattices.
\end{abstract}


\maketitle

{\sc Introduction --} Parity (${\cal P}$) and time -reversal (${\cal T}$) symmetries, as well as their 
breaking, belong to the most basic notions in physics. Recently there has been much interest in 
systems which do not obey ${\cal P}$ and ${\cal T}$-symmetries separately but do exhibit a combined 
${\cal PT}$-symmetry. Examples of such ${\cal PT}$-symmetric systems range from quantum field theories 
to solid state and classical optics \cite{Bender2007, Bender1999,Mus2008a,BB98,GGKN08,Makris2008,GMCM07,
RDM05,B08}. A $\mathcal{PT}$-symmetric system can be realized in optics, by creating a medium with 
alternating regions of gain and loss, such that the (complex) refraction index satisfies the condition 
$n^*(-x)=n(x)$ \cite{Makris2008,GMCM07,RDM05,B08}. This condition implies that creation and absorption 
of photons occur in a balanced manner, so that the net loss or gain is zero. 
Such synthetic ${\cal PT}$ metamaterials show unique characteristics such as ``double refraction" and 
non-reciprocal diffraction patterns, which may allow their use as a new generation of unidirectional 
optical couplers or left-right sensors of propagating light \cite{Makris2008}. In the paraxial approximation 
the classical wave equations reduces to a Schr\"{o}dinger equation with a fictitious time, related to 
the propagation distance, and with the refraction index playing the role of the potential. We use below
the terminology of the Schr\"{o}dinger equation, while keeping in mind applications to optical systems.

A PT-symmetric system can be described by a phenomenological "Hamiltonian" ${\cal H}$. Such Hamiltonians 
may have a unitary time evolution and a real energy spectrum, although in general are non-hermitian. 
Furthermore, as some parameter of ${\cal H}$ changes, a spontaneous ${\cal PT}$ 
symmetry breaking occurs, at which point the eigenfunctions of ${\cal H}$ cease to be eigenfunctions of 
the ${\cal PT}$-operator, despite the fact that ${\cal H}$ and the $\mathcal{PT}$-operator commute 
\cite{Bender2007}. This happens because the ${\cal PT}$-operator is not linear, and thus the eigenstates 
of ${\cal H}$ may or may not be eigenstates of ${\cal PT}$. As a consequence, in the {\it broken $\cal{PT}$
-symmetry phase} the spectrum becomes partially or completely complex. The other limiting case where both 
${\cal H}$ and ${\cal PT}$ share the same set of eigenvectors, corresponds to the so-called {\it exact 
$\mathcal{PT}$-symmetric phase} and the spectrum is real.

In this Letter we investigate the spontaneous $\mathcal{PT}$-symmetry breaking scenario in a wide class 
of systems supporting localized states. Such states are ubiquitous in macroscopic systems. They can reside 
on impurities or at the edges of an otherwise perfect lattice of finite size. Therefore, in order to 
understand the $\mathcal{PT}$-symmetric phase of a macroscopic system, it is imperative to consider localized 
states. At the same time, we note that even fifty years after the seminal work 
of Anderson \cite{A58}, localization continues to be a thriving area of research, not only for solid-state 
physics, but also to other fields including ultra-cold atoms, acoustics, 
microwaves and classical optics. We therefore expect that our study linking the newly developed area of 
$\mathcal{PT}$ materials with the field of localization will contribute to understanding fundamental 
aspects of modern physics.

Localization is particularly pronounced in one-dimensional (1D) systems and has been studied 
extensively in the past \cite{Lifshits1988}. We show that in case of ${\cal PT}$-symmetric lattices of size 
$N$ which can support localized modes due to disorder or impurities or even due to boundaries (surface states), 
the mechanism that triggers the transition from real to complex spectrum is level crossing between a 
pair of modes having the smallest energy spacing. Due to the ${\cal P}$-symmetry, this pair of states has a 
double hump shape and the energy splitting between them is $\delta_1 \sim \exp (-l_0/\xi)$ where $\xi$ is the 
localization length and $l_0$ is the distance between the two humps (for disordered lattices $l_0\sim N$). 
We find that the value of the ${\cal PT}$ symmetry breaking parameter $\gamma$ at the transition point is 
$\gamma_{\cal PT}\sim \delta_1$, thus indicating that {\it the exact ${\cal PT}$-symmetric phase is exponentially 
small} in the limit $l_0/\xi \gg 1$. In contrast, for $l_0/\xi\ll 1$, we find that the smaller level spacing 
scales as $\Delta_{\rm min}\sim 1/N^2$. This is also reflected in the distribution ${\cal P}(\gamma_{
\cal PT})$ which changes from a log-normal towards a semi-Gaussian as $N/\xi$ decreases.

{\sc Two PT-symmetric Impurities --}
It is instructive to start with the simple example of a pair of $\mathcal{PT}$-symmetric impurities implanted 
into an otherwise perfect infinite lattice. The system is described by the equation
\begin{eqnarray}
-\psi_{n+1}-\psi_{n-1} =(E-\varepsilon_{n})\psi_{n}
\label{eq1}\,,
\end{eqnarray}
where $\psi_n$ is the eigenfunction amplitude at site $n$, $\varepsilon_{n}=0$ for $n\neq\pm l$, and 
$\varepsilon_{\pm l}=-\beta\pm i\gamma$, with $\beta$ and $\gamma$ being real and positive. 
We are looking for the bound states:
\begin{equation}
\psi_{n} = \left\{
  \begin{array}{lr}
    A e^{kn}, & n\le -l\\
    B e^{kn} + C e^{-kn}, & -l\le n\le l\\
    D e^{-kn}, & n\ge l
  \end{array}
\right.
\label{eq3}
\end{equation}
with $\text{Re}[k]>0$ and $E=-2\cosh k$. Matching the wave function at the sites $n=\pm l$, one obtains 
four equations for the amplitudes $A,B,C,D$. Equating the determinant to zero yields the transcendental 
equation for $k$:
\begin{eqnarray}
\sinh k=\frac{\beta}{2}\pm\frac{1}{2}\left[ -\gamma^{2} +
        (\beta^{2}+\gamma^{2})e^{-4kl} \right]^{1/2} \label{eq4}\,.
\end{eqnarray}
For $\gamma=0$ and $\beta l\gg 1$, one finds two bound states with energies $E_{\pm}=E_{0}\pm \frac{1}
{2} \delta_1$, where $E_{0}=-\sqrt{4+\beta^{2}}$ is the energy of a localized state on a single isolated 
impurity, and $\delta_1 = \delta (l) = (2\beta^{2}/\left|E_{0}\right|)e^{-\beta l}$ is the exponentially 
small energy splitting term for the two-impurity problem. The point we want to emphasize is that for 
$\beta l\gg 1$ 
already an exponentially small $\gamma$ leads to complex values of $k$ and $E$, thus, breaking the 
${\cal PT}$-symmetry. The mechanism for this breaking is level crossing: as follows from Eq. (\ref{eq4}), 
when $\gamma$ reaches the value $\gamma_{\cal PT}\approx\beta e^{-\beta l}$, the two (real) eigenvalues 
become degenerate. For $\gamma> \gamma_{\cal PT}$ they branch out into the complex plane, displaying 
near the branch point the characteristic behavior $\text{Im}[E_{\pm}]\varpropto\pm\sqrt{\gamma^{2}-
\gamma_{\cal PT}^{2}}$. This square root singularity seems to be a generic feature of the ${\cal PT}$
-symmetry breaking.

The eigenfunctions of the above Hamiltonian also undergo characteristic changes as $\gamma$ 
increases. A finite $\gamma$ breaks the $\mathcal{P}$-symmetry of the Hamiltonian but, as long as 
$\gamma < \gamma_{\cal PT}$, the $\mathcal{PT}$-symmetry of the eigenfunction is preserved, so that 
$\psi^{\ast}_{n} = \pm\psi_{-n}$. This implies that the ``dipole moment'', $D=\sum_{n=-N}^{N} n\left|\psi_{n}
\right|^{2}$, of an eigenstate is zero. For $\gamma > \gamma_{\cal PT}$ the eigenstates acquire a 
finite dipole moment. Below we shall use $D$ as one of the signatures of the ${\cal PT}$-symmetry
breaking.

\begin{figure}
\includegraphics[width=\linewidth]{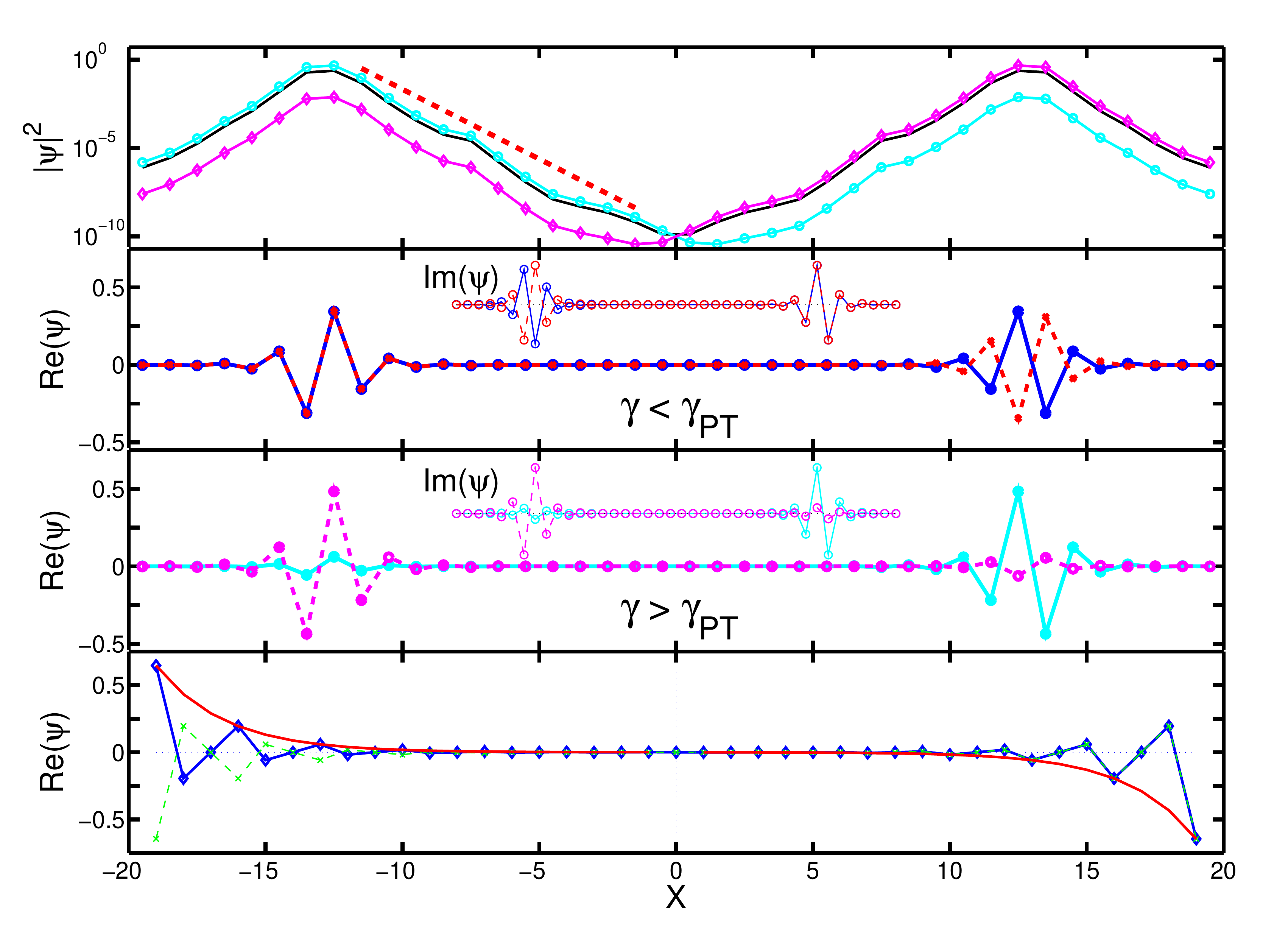}
\caption{Pairs of exponentially localized modes in a 1D-chain with ${\cal PT}$-symmetric disorder 
(a-c) and surface states in a ${\cal PT}$-symmetric periodic chain (d) yielding the smallest energy splittings 
$\delta_1$ for $\beta=3$. For $\gamma<\gamma_{\cal PT}$ (b) the two eigenfunctions 
(blue and red; the imaginary part is shown in the inset) are complex but ${\cal PT}$ symmetric and their 
absolute values remains equal and symmetric [coinciding on the black line in (a)]. For $\gamma>\gamma_{
\cal PT}$ (c) the eigenfunctions (magenta and cyan) are no longer ${\cal PT}$-symmetric and shift their 
weight towards separate sides of the chain (a). (d) Surface states (blue and green) showing exponential
localization. The red dashed [solid] lines in (a) [(d)] are guides to the eye, pointing out the exponential 
localization. 
}
\end{figure}

{\sc The disordered PT-symmetric Chain --}
We consider next a 1D disordered $\mathcal{PT}$-symmetric chain and demonstrate that under a disorder
increase, the $\mathcal{PT}$-symmetric phase is gradually destroyed. For sufficiently strong disorder 
this phase shrinks to an exponentially narrow region, even for a comparatively small system.

The chain is described by the Hamiltonian of Eq. (\ref{eq1}), where now all $\varepsilon_{n}$ are 
random complex numbers, $\varepsilon_{n}=\beta_{n}+i\gamma_{n}$, with the constraint $\varepsilon_{n}
=\varepsilon^{\ast}_{-n}$. One can envisage various possibilities for randomness, either in $\beta_n$ 
or $\gamma_n$, or both. Below we present results for the case where $\beta_{n}$ (for $n\ge0$) are 
uniformly distributed on the interval $[-\beta/2;\beta/2]$ and $\gamma_{n}=\gamma =const.$ (for $n\ge1$).
It is crucial, for the $\mathcal{PT}$-symmetry, to implement the constraint $\beta_{n}=\beta_{-n}$ and 
$\gamma_{n}=-\gamma_{-n}$ (the latter implies $\gamma_{0}=0$). This constraint introduces a peculiar 
long-range correlation. To clarify the picture we start with the Hermitian limit $\gamma=0$, 
and assume a long chain, such that even eigenstates in the band center are localized, i.e. their 
localization length is smaller than the system size ($2N+1$). Imagine for a moment that the chain is cut 
in half, at its center $n=0$. Then a typical state in one half of the chain would be localized, with some 
localization length $\xi$, far away from $n=0$, say, near site 
$l\gg1$. This state has its counterpart in the other half of the chain, near site $-l$. In the full, 
connected chain this pair of states has an exponentially small overlap, at the center of the chain, 
yielding two real eigenstates of the entire chain. Each of these eigenstates (one symmetric, the 
other antisymmetric) has two peaks, near the sites $n=\pm l$. The energy splitting between the two 
eigenvalues is exponentially small, $\delta (l) \simeq e^{-2l/\xi}$, in complete analogy with the 
example of the two impurities in a perfect chain.

Thus, the eigenstates in a $\mathcal{P}$-symmetric disordered chain are organized into pairs (doublets)
of energy splitting $\delta_1< \delta_2<\cdots$. The energy splitting between the two states 
of a doublet is exponentially small, while the energy separation between consecutive doublets is 
much larger, of the order of level spacing, $\Delta$, in half chain. The pair associated with the 
minimum splitting, $\delta_{1}$ is likely to originate from states localized far away from the origin 
of the chain ($n=0$) and with energies at the band edges (small $\xi$). As $\gamma$ is switched on 
the eigenstates of each pair will initially preserve their ${\cal PT}$-symmetric structure (see Fig. 
1a,b). At $\gamma =\gamma_{\cal PT} \simeq \delta_{1}$, the two levels associated with $\delta_{1}$ 
will cross, breaking the $\mathcal{PT}$-symmetry (see Fig. 2a). As $\gamma>\gamma_{\cal PT}$ these
modes cease to be eigenstates of the ${\cal PT}$-operator. Instead, the weight of each of them is 
gradually shifted towards one of the localization centers. An example of such pair associated with 
$\delta_1$ is reported in Fig. 1a,c. 
For larger $\gamma$ the next doublet, with splitting $\delta_2$ 
will come into play, creating a second pair of complex eigenvalues for $\gamma 
\simeq \delta_2$ (see Fig. 2a), etc.

The above qualitative considerations apply to a single, realization of the random potential. A full 
theory must be formulated in statistical terms and deal with probability distributions. For instance, 
the critical value $\gamma_{\cal PT}$ at which the ${\cal PT}$ -symmetry is broken, fluctuates from 
one realization to another and the appropriate question pertains to the distribution $P(\gamma_{\cal 
PT})$. As has been argued above (see inset of Fig. 2a), 
in the strong disorder regime $\gamma_{\cal PT}\approx \delta_{1}$, and thus the problem reduces to 
the study of the distribution ${\cal P}(\delta_{1})$. There are several sources of fluctuations 
in $\delta_{1}$: fluctuations in the position and energy of the relevant localized states, as well as 
what can be termed ``fluctuations in the wave functions''. By this we mean that a localized wave function 
exhibits strong, log-normal fluctuations in its ``tails'', i.e. sufficiently far from its localization 
center \cite{Altshuler1989}. This latter source of fluctuations appears to be the dominant one and it 
immediately yields a log-normal distribution for $\delta_{1}$ (see Fig. 2b), since $\delta_1$ is 
proportional to the overlap integral between a pair of widely separated and strongly localized states.

\begin{figure}
\includegraphics[width=\linewidth]{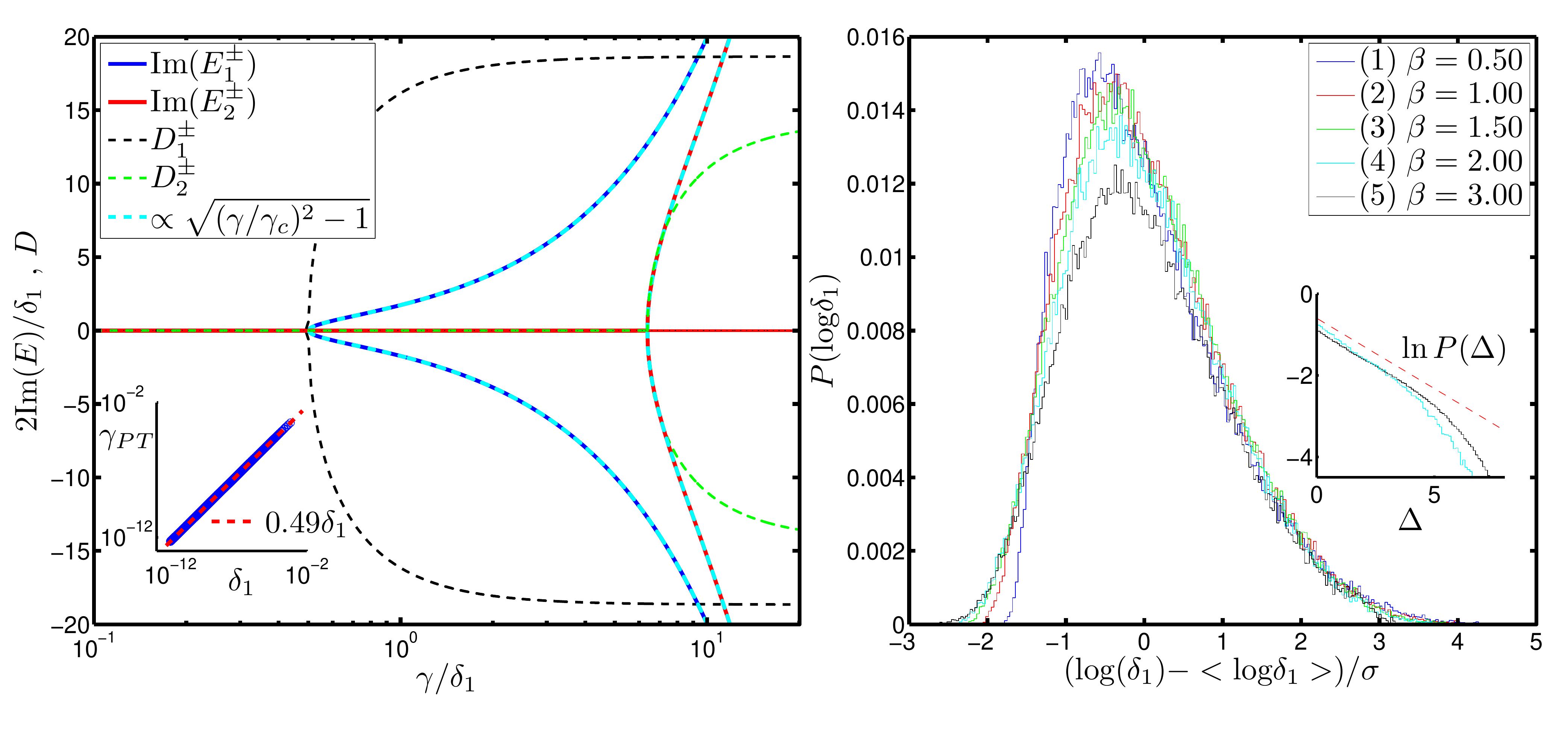}
\caption{
(a) Bifurcations for the dipole moment $D$ and for the imaginary part ${\rm Im} E$ of energy levels 
for a ${\cal PT}$-symmetric 1D chain with $\gamma=$const. and $\beta_n$ given by a box distribution 
$[-{\beta\over 2};{\beta \over 2}]$ for $N/\xi\gg 1$. The first two bifurcations (corresponding to 
splittings $\delta_1$ and $\delta_2$) are shown. The square-root behavior at the branching point (see 
text) is indicated with dashed cyan lines. Inset: a linear relation between $\delta_1$ and $\gamma_{
\cal PT}$ is evident for almost $10$ orders of magnitude. (b) Distribution ${\cal P}(x)$ of $x\equiv
(\log\delta_1-\langle\log\delta_{1}\rangle)/\sigma$ ($\sigma$ is the standard deviation) for various 
disorder strengths $\beta$. In the limit of large $\beta$ the distribution becomes log-normal. Inset: 
The distribution ${\cal P}(s)$ of $s\equiv \log(\delta_2)-\log(\delta_1)$ reported in a 
semi-logarithmic plot. A Poisson distribution is approached as $\beta$ increases.
}
\end{figure}

The aforementioned scenario of bifurcations, i.e. of the consecutive branching of pairs of eigenvalues 
into the complex plane, can also be made more quantitative. Consider, the separation (on 
the $\gamma$-axis) between the first bifurcation ($\gamma=\gamma_{\cal PT}$) and the next one. This 
separation, $\delta\gamma$, is proportional to $(\delta_2-\delta_{1})$. Assuming that localized states, 
associated with the doublets, are randomly located, and ignoring for the moment fluctuations in the 
energy of these states, one immediately obtains a Poisson distribution for $s\equiv\log\delta_2-
\log\delta_{1}$, with the average spacing $2/\xi$ on the $\log\gamma$ scale between the bifurcation 
points (see inset of Fig. 2b). This result, with $\xi$ being replaced by an appropriate average, remains 
valid also when we account for the energy fluctuations. 

Our considerations can be extended to the $N/\xi \ll 1$ limit, when the states responsible for $\delta_{1}$
are extended over the entire chain. In this case the picture 
of doublets with exponentially small energy splittings is not valid and $\gamma_{\cal PT}$ becomes of 
the order of the minimal level spacing, $\Delta_{\rm min}$, in the corresponding 
Hermitian problem. This statement follows from perturbation theory, with respect to $\gamma$. The 
unperturbed (i.e. $\gamma=0$) energy levels are real, and are separated by intervals of 
order $1/N^2$ at the band-edges (at the center of the band the separation is of order $1/N$), so that
$\Delta_{\rm min}\simeq 1/N^2$. Finite $\gamma$ leads to level shifts proportional to $\gamma^2$ (the first 
order correction vanishes due to ${\cal PT}$-symmetry) and for $\gamma=\gamma_{\cal PT}\simeq\Delta_{\rm min}$ 
the perturbation theory breaks down, signaling level crossing and the appearance of the first pair of 
complex eigenvalues. Thus, the energy scale for the ${\cal PT}$-threshold in the $N/\xi \ll 1$ limit 
($\gamma_{\cal PT}\simeq 1/N^2$) widely differs from that for $N/\xi \gg 1$ ($\gamma_{\cal PT}
\simeq e^{-2N/\xi}$). However, the ``bifurcation scenario'', with characteristic square-root branches 
for the complex eigenvalues, holds in both cases (again, with the completely different energy scale for 
the intervals between the consecutive bifurcations). Our numerical results presented in Fig. 3a are in
perfect agreement with these considerations.

We study now the distribution $P(\gamma_{\cal PT})$ in the limit $N/\xi\ll 1$. We invoke perturbation 
theory with respect to the perfect lattice. The perturbative scenario, indicates 
that weak disorder will cause a small shift of the energy levels. Thus the new level spacing becomes 
$\Delta_{\rm min}\pm\Delta\delta$, where the sign $+$ ($-$) refers to the upper (lower) band-edge. 
Regardless of the sign of $\Delta\delta$, the minimal level spacing (in first order perturbation theory)
is
\begin{equation}
\delta = \Delta_{\rm min}-\left|\Delta\delta\right|\equiv
\Delta_{\rm min}-{4\over 2N+1}\left|\sum\limits_{n=1}^{N} A_{n}\beta_{n}\right| ,
\end{equation}
with the coefficients $A_{n}=\sin\left(\frac{\pi (n+N)}{2N+1}\right) \sin\left(\frac{3\pi (n+N)} {2N+1}
\right)$. 
If $\beta_{n}$ were Gaussian distributed, it would be immediately clear that the distribution of $\delta$, 
$P(\delta)$, is a semi-Gaussian for $\delta<\Delta_{\rm min}$. This should remain approximately true also 
for the box-distribution, employed in our numerics, if the number of terms in the sum is sufficiently large. 
Fig 3b confirms this expectation.

{\sc Periodic PT-symmetric Potentials --}
Let us briefly address the problem of $\mathcal{PT}$-symmetry breaking for a periodic complex potential.
This question has been raised in \cite{Makris2008}, for a potential $V(x) = 4(\cos^{2} x + iV_{0} \sin 2x)$, 
where it was argued that the critical value, of the $\mathcal{PT}$-threshold was $V^{\text{th}}_{0}=1/2$. 
The presence of the real part, 
$A\cos^{2} x$ ($A\ne0$), is crucial for this result. Indeed, it was pointed out in \cite{Makris2008} 
that a purely imaginary periodic potential, $V(x)=iV_{0}\sin^{2N+1}(x)$, treated in Ref \cite{Bender1999}, 
has ``zero $\mathcal{PT}$-threshold'', i.e. it cannot have an entirely real spectrum. Another example of 
a periodic potential with zero $\mathcal{PT}$-threshold was provided in Ref \cite{Mus2008a}.

\begin{figure}
\includegraphics[width=\linewidth]{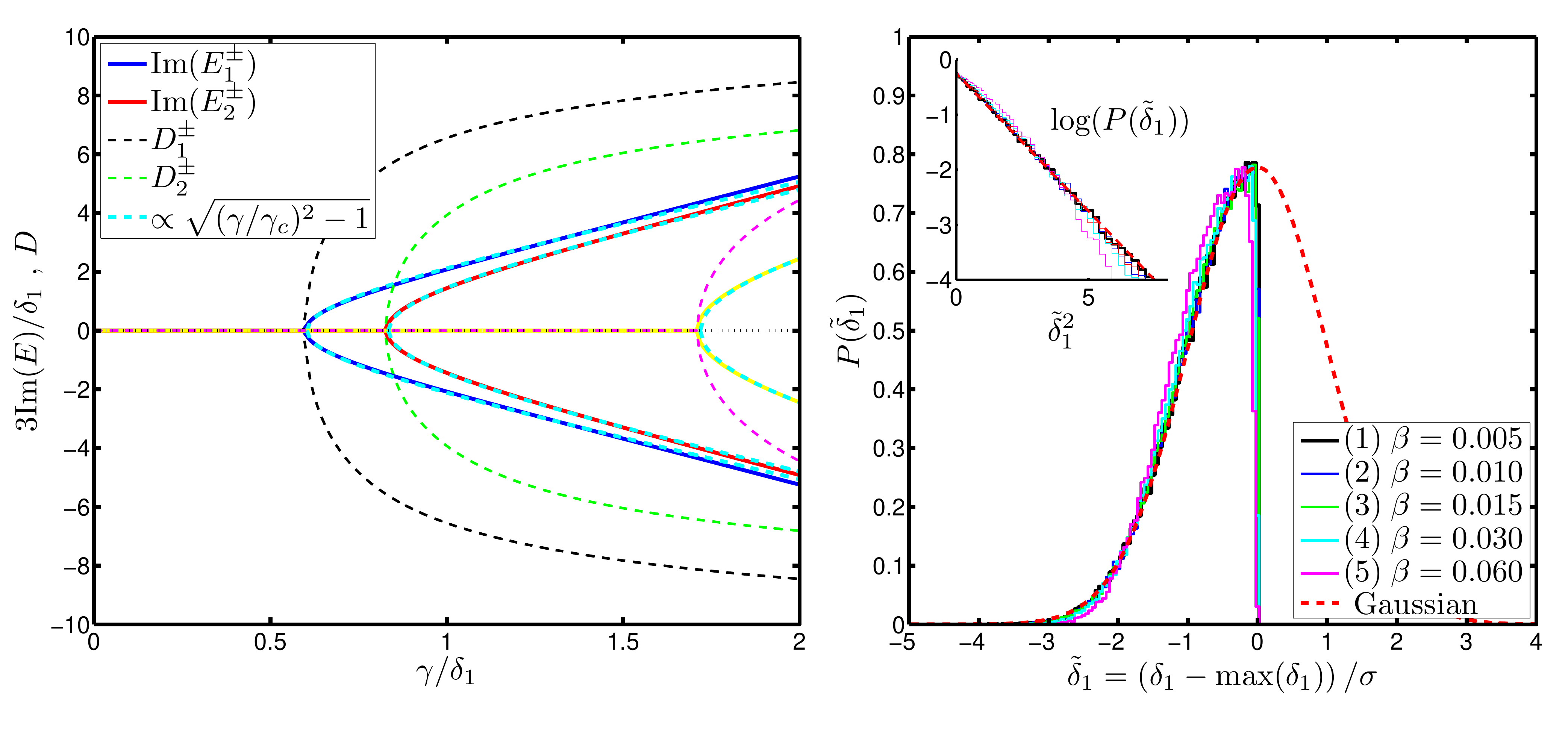}
\caption{(a) Same as in Fig. 2a but now for "weak" disorder $N/\xi \ll 1$. The same bifurcation scenario 
is observed. (b) The distribution ${\cal P}({\tilde \Delta})$ of the rescaled minimal energy split $
{\tilde \Delta}=\Delta \delta/\sigma$ ($\sigma$ is the standard deviation), for various disordered 
strengths, all of them being in the weak localized regime. The distribution has an upper cutoff at 
${\tilde \Delta}=0$. A Gaussian distribution is shown also by a red dashed line. Inset: The same data 
in a semi-logarithmic manner vs. ${\tilde \Delta}^2$.
}
\end{figure}

In a periodic system of finite extent one usually encounters localized states (surface states) at the
boundaries of the sample \cite{Tamm1932}. We have found that these states, which were not addressed in 
Ref. \cite{Makris2008}, are crucial for the correct evaluation of the ${\cal PT}$-threshold in a finite
system. We illustrate the importance of the surface states by taking the example of a tight binding 
periodic potential, with three sites per unit cell. The Hamiltonian is still that of Eq.(\ref{eq1}), 
but with on-site energies: $\varepsilon_{n=3l}=0; \varepsilon_{n=3l\pm 1} = \beta \pm i\gamma$ 
where $l=0,\pm1,\pm2,\ldots,\pm N/3$. For $\gamma=0$ (and $N \rightarrow\infty$) the spectrum displays 
two energy gaps, whose width (for $\beta\ll1$) is $2\beta/3$. The existence of gaps suggests, in analogy 
with \cite{Makris2008}, that the 
$\mathcal{PT}$-symmetric phase in this model will exhibit some robustness, as long as $\gamma$ is small 
($\gamma\ll\beta$). This expectation, however, is not born out due to the surface states. For 
$\gamma=0$ there is a pair of surface states, exponentially decaying away from the sites $\pm N$.
For large but finite $N$ these surface states have a small overlap near the center of the chain $n=0$, 
yielding a doublet with an exponentially small energy splitting. In complete analogy with the two-impurity 
problem, already an exponentially small $\gamma\simeq e^{-\beta N}$ will therefore break 
the $\mathcal{PT}$-symmetry. This example shows that the $N\rightarrow\infty$ limit can be quite subtle, 
as far as the $\mathcal{PT}$-threshold is concerned. If one starts directly with $N=\infty$, one obtains 
a finite $\mathcal{PT}$-threshold, $\gamma_{\cal PT}\simeq\beta$. If one starts, however, with a large 
but finite $N$ and then takes the $N\rightarrow\infty$ limit (which is the correct 
physical limit), then one ends up with $\gamma_{\cal PT}=0$, due to the existence of surface states.

{\sc Conclusion --}
We have studied a 1D $\mathcal{PT}$-symmetric chain with disorder. The $\mathcal{PT}$-
symmetric phase turns out to be very fragile. For a sufficiently long chain, this phase exists only 
for exponentially small values of the imaginary part of the potential $\gamma_{\cal PT}
\simeq e^{-N/\xi}$ beyond which the $\mathcal{PT}$-symmetry is broken (here $N$ and $\xi$ are the system 
size and the localization length respectively). Our model can be extended in various directions. For 
instance, we have checked that our main results do not change if randomness is introduced in both
the real and the imaginary parts of the potential. We have also briefly discussed the 
periodic $\mathcal{PT}$-potential and pointed out the importance of surface states in breaking ${\cal 
PT}$-symmetry. Our main conclusion is within a generic 1D system which supports localized modes, the 
threshold for $\mathcal{PT}$-symmetry breaking is exponentially approaching zero with increasing size.

\begin{acknowledgments}
We acknowledge correspondence with K.~Makris and D.~Christodoulides concerning the 
zero $\mathcal{PT}$-threshold for periodic potentials considered in Ref \cite{Makris2008} and V. Kovanis
for attracting our interest to ${\cal PT}$-systems. This work has been supported by a grant 
from the US-Israel Binational Science Foundation (BSF) and the DFG FOR760. 
\end{acknowledgments}


\end{document}